\documentclass[amsmath,amssymb,prc,final,showpacs,twocolumn]{revtex4-1} 

\usepackage{bm,hyphenat,xspace} 
\usepackage{graphicx,epsfig}

\newcommand{\A}[2]{{}^{#1}\mathrm{#2}}

\newcommand {\mbf}[1]{{\mathbf{#1}}}
\newcommand {\mct}{{\mathcal{T}}}
\newcommand {\vecg}[1]{\mbox{\boldmath{$#1$}} }
\newcommand{\cm}{\mathrm{c\!\:\!.m\!\:\!.}}

\begin{document}

\title {Faddeev-type calculation of $(d,n)$ transfer reactions in 
three-body nuclear systems}

\author{A.~Deltuva} 
\email{arnoldas.deltuva@tfai.vu.lt}
\affiliation{Institute of Theoretical Physics and Astronomy, 
Vilnius University, A. Go\v{s}tauto 12, LT-01108 Vilnius, Lithuania}

\received{October 11, 2015}

\pacs{24.10.-i, 21.45.-v, 25.45.Hi, 25.40.Hs}

\begin{abstract}
Exact Faddeev-type three-body equations are applied to the
study of the proton transfer  reactions  $(d,n)$  in the system consisting of
a nuclear core and two nucleons. The integral equations for the
three-body transition operators are solved in the momentum-space 
framework including the Coulomb interaction via the screening
and renormalization method. For a weakly bound final nucleus
the calculation of the $(d,n)$ reaction is more demanding in terms
of the screening radius as compared to the $(d,p)$ reaction.
Well converged differential cross section
results are obtained for $\A{7}{Be}(d,n)\A{8}{B}$,
$\A{12}{C}(d,n)\A{13}{N}$, and  $\A{16}{O}(d,n)\A{17}{F}$ reactions.
A comparison with the corresponding $(d,p)$ reactions is made.
The calculations fail to reproduce the shape of the angular distribution
for reactions on $\A{12}{C}$ but
provide quite successful description for   reactions on $\A{16}{O}$,
especially for the transfer to  the $\A{17}{F}$ excited state $1/2^+$
when using a nonlocal optical potential.
\end{abstract}

 \maketitle

\section{Introduction \label{sec:intro}}

Although there is a long history of nuclear reaction calculations using
three-body models  \cite{johnson:70a,austern:87}, 
the rigorous Faddeev-type scattering theory 
\cite{faddeev:60a,alt:67a} was practically
applied to nuclear reaction problem only in the last decade
\cite{deltuva:07d,deltuva:09a}. 
Despite being theoretically most complicated and computationally
more expensive than traditional approximate three-body reactions methods,
the Faddeev formalism has an advantage that, once numerically 
well-converged results are obtained, the discrepancies with the experimental data can be 
attributed  to the shortcomings of the used  potentials
or to the inadequacy of the three-body model.
The numerical calculations \cite{deltuva:07d,deltuva:09a} have been performed using
the Alt-Grassberger-Sandhas (AGS) integral equations for transition operators
\cite{alt:67a} that were solved in the momentum-space framework;
the Coulomb interaction was included via the screening and renormalization
method \cite{taylor:74a,alt:80a,deltuva:05a}. 
So far the applications of Faddeev/AGS equations are limited to three-body systems made of
a proton ($p$), neutron ($n$), and  nuclear core ($A$).
With $A$ most often being one of $\A{10}{Be}$, $\A{12}{C}$,  $\A{14}{C}$, or  $\A{16}{O}$, 
reactions initiated by the collisions of the deuteron ($d$) with the nucleus $A$ and of the proton
with the bound system $(An)$ have been studied, 
including the elastic proton and deuteron scattering,
i.e., $(p,p)$ and $(d,d)$ processes, deuteron breakup $(d,pn)$ and one-neutron
removal $(p,pn)$, deuteron stripping $(d,p)$ and pickup $(p,d)$,
and, to a lesser extent, the charge-exchange reaction $(p,n)$.
However, to date there are no deuteron stripping $(d,n)$ and its time-reverse $(n,d)$
reaction calculations in three-body  Faddeev/AGS equation framework.
Among the two the $(d,n)$ process is especially important since it may be used for the creation
 and study of weakly bound core plus valence proton $(Ap)$ systems such as one proton
halo nucleus  $\A{8}{B}$.
Therefore the aim of the present work is to extend the rigorous three-body  Faddeev/AGS framework
and to apply it to the study of $(d,n)$  reactions.

In Sec.~\ref{sec:2} the Faddeev/AGS formalism is recalled and specific aspects of 
$(d,n)$  reaction calculations are pointed out. 
In Sec.~\ref{sec:3} physics results are presented for the $(d,n)$ reactions
on  $\A{7}{Be}$, $\A{12}{C}$, and  $\A{16}{O}$ nuclei; for the latter two the 
comparison with the corresponding $(d,p)$ reactions is made as well;
the results for the elastic scattering, extensively studied in previous works
\cite{deltuva:07d,deltuva:09a,deltuva:09b}, are not shown.
The summary is given in Sec.~\ref{sec:4}.

\section{Theoretical framework \label{sec:2}}

The AGS formalism is an integral equation formulation of the exact three-body scattering
theory. Instead of the wave function it deals with transition operators that
contain  the full physical information about the considered process.
The AGS integral equations are most convenient to solve in the momentum-space representation.
The standard AGS formalism assumes short-range potentials within the three pairs of
particles. This condition is fulfilled by the nuclear interactions
$v_A$, $v_p$, and $v_n$ that, in the odd-man-out notation, denote
$p$-$n$, $n$-$A$, and $A$-$p$ potentials, respectively. However, the
proton-core Coulomb repulsion $w_n$, in the coordinate space given as
 $w_n(r) = Z \alpha_{e}/r$ with $Z$ being the charge number of nucleus $A$ and
$\alpha_e \approx 1/137$ the fine structure constant, is of the long range.
Nevertheless,  $w_n$  can be included rigorously using
the screening and renormalization
method \cite{taylor:74a,alt:80a,deltuva:05a}. 
For this purpose the screened Coulomb potential
\begin{equation}  \label{eq:wr}
w_{nR}(r) = w_n(r) e^{-(r/R)^{\bar{n}}},
\end{equation}
where $R$ is the screening radius and $\bar{n}$ is the screening smoothness parameter,
is added to the nuclear one allowing the standard scattering theory to be applied for the sum
$v_n + w_{nR}$.

Via the Lippmann-Schwinger integral equation the  pair potentials yield the
corresponding two-particle transition operators 
\begin{subequations}  \label{eq:LS}   
\begin{align}  
T_{A} = {}& v_A +  v_{A} G_0 T_A, \\
T_{p} = {}& v_p +  v_{p} G_0 T_p, \\
T_{n}^{(R)} = {}& v_n+w_{nR} +  (v_n+w_{nR}) G_0 T_{n}^{(R)}.
\end{align}
\end{subequations}
Here  $G_0 = (E+i0 -H_0)^{-1}$ is the free resolvent at the energy $E$ available
for the relative three-body motion, and $H_0$ is the respective three-body
kinetic energy operator. The two-body transition operators, when iterated to
all orders via AGS equations, lead to the three-body transition operators. The 
deuteron-nucleus scattering process is described by the set
\begin{subequations}  \label{eq:AGS}   
\begin{align}  
U_{AA}^{(R)} = {}& T_p G_0 U_{pA}^{(R)} +   T_n^{(R)} G_0 U_{nA}^{(R)}, \\
U_{pA}^{(R)} = {}& G_0^{-1}  +   T_n^{(R)} G_0 U_{nA}^{(R)} + T_A G_0 U_{AA}^{(R)} , \\
U_{nA}^{(R)} = {}& G_0^{-1}  + T_A G_0 U_{AA}^{(R)} + T_p G_0 U_{pA}^{(R)}.
\end{align}
\end{subequations}
Through $T_{n}^{(R)}$ also the three-body transition operators acquire
the dependence on the $A$-$p$ Coulomb screening radius $R$.
The transition amplitudes for the two-cluster reactions initiated by 
the $d$+$A$ collisions are determined by the on-shell matrix elements
of  three-body transition operators between the respective initial and 
final two-cluster channel states 
$|\Phi_A(\mbf{q}_A)\rangle = |\phi_A\rangle |\mbf{q}_A\rangle$,
$|\Phi_p(\mbf{q}_p)\rangle = |\phi_p\rangle |\mbf{q}_p\rangle$, and
$|\Phi_n^{(R)}(\mbf{q}_n)\rangle = |\phi_n^{(R)}\rangle |\mbf{q}_n\rangle$.
Here $|\mbf{q}_\alpha\rangle$ is a free wave for the $\alpha$ spectator-pair
relative motion with the momentum $\mbf{q}_\alpha$ while
$|\phi_A\rangle$, $|\phi_p\rangle$, and  $|\phi_n^{(R)}\rangle$ are two-body
bound state wave functions for the $(pn)$, $(nA)$, and $(Ap)$ subsystems
calculated with the potentials $v_A$, $v_p$, and $v_n+w_{nR}$, respectively.
The dependence on the other discrete quantum numbers is suppressed in the notation.
None of the matrix elements
$\langle \Phi_A(\mbf{q}'_A)|U_{AA}^{(R)}|\Phi_A(\mbf{q}_A)\rangle $,
$\langle \Phi_p(\mbf{q}_p)|U_{pA}^{(R)}|\Phi_A(\mbf{q}_A)\rangle $,
$\langle \Phi_n^{(R)}(\mbf{q}_n)|U_{nA}^{(R)}|\Phi_A(\mbf{q}_A)\rangle $ has the
$R\to \infty$ limit, however, after the renormalization with the appropriate
(diverging as well) phase factors $Z_{\alpha R}$
the infinite $R$ limit exists \cite{taylor:74a,alt:80a,deltuva:05a} and 
corresponds to the physical transition amplitudes 
\begin{subequations}  \label{eq:UZ}   
\begin{align}  
\nonumber
\mct_{AA}(\mbf{q}'_A,\mbf{q}_A) = {}& t_{A}^{C}(\mbf{q}'_A,\mbf{q}_A) 
 + \lim_{R\to \infty} \big[ Z_{AR}^{-\frac12}(q'_A) \langle \Phi_A(\mbf{q}'_A)| \\ {}& \times
(U_{AA}^{(R)}- t_{A}^{R})|\Phi_A(\mbf{q}_A)\rangle Z_{AR}^{-\frac12}(q_A) \big],
 \\ \nonumber
\mct_{pA}(\mbf{q}_p,\mbf{q}_A) = {}& \lim_{R\to \infty} \big[
 Z_{pR}^{-\frac12}(q_p)
\langle \Phi_p(\mbf{q}_p)| \\ {}& \times
U_{pA}^{(R)}|\Phi_A(\mbf{q}_A)\rangle  Z_{AR}^{-\frac12}(q_A) \big], \\ \label{eq:UZn}  
\mct_{nA}(\mbf{q}_n,\mbf{q}_A) = {}& \lim_{R\to \infty} \big[
\langle \Phi_n^{(R)}(\mbf{q}_n)|U_{nA}^{(R)}|\Phi_A(\mbf{q}_A)\rangle Z_{AR}^{-\frac12}(q_A) \big].
\end{align}
\end{subequations}
In the case of the elastic  scattering the longest-range screened Coulomb contribution
$t_{A}^{R}$, corresponding to the Coulomb interaction between the nucleus and the center-of-mass (c.m.)
of the deuteron, is separated from $U_{AA}^{(R)}$ and renormalized analytically
in the infinite $R$ limit, leading to the standard Rutherford amplitude 
$t_{A}^{C}(\mbf{q}'_A,\mbf{q}_A) $ \cite{taylor:74a}.
 All remaining terms have to be calculated numerically,
but  owing to their short-range nature, the convergence with $R$ is quite fast as will be demonstrated.
Since the AGS equations are solved in the partial-wave representation, the renormalization
factors are most conveniently calculated as
\begin{equation}  \label{eq:zr}
Z_{\alpha R}(q_\alpha) = e^{-2i [ \sigma_{\alpha, L}^C(q_\alpha)- \eta_{\alpha, L}^R(q_\alpha)]},
\end{equation}
where the full and screened Coulomb phase shifts $\sigma_{\alpha, L}^C(q_\alpha)$
and $ \eta_{\alpha, L}^R(q_\alpha)$ correspond to the relative motion with the angular momentum
$L$ between the particle $\alpha$ and  c.m. of the remaining pair.

Equation \eqref{eq:UZn} takes into account that there is no Coulomb interaction between
the neutron and the $(Ap)$ pair, i.e., $Z_{n R}(q_n) =1$ and the final
$n+(Ap)$ state is not distorted by the Coulomb force. However, in contrast to other channel states,
 the bound state  $|\phi_n^{(R)}\rangle$ is affected by the screened Coulomb interaction.
Due to $|\phi_n^{(R)}\rangle$ one may expect for observables of the $(d,n)$ reaction  a different 
convergence rate with increasing $R$ as compared to the $(d,p)$ reaction.
These differences will be studied in the next section.

\section{Results \label{sec:3}}

As already mentioned, the AGS integral equations \eqref{eq:AGS} 
are solved numerically in the momentum-space partial-wave basis.
Typically, the potentials  $v_A$, $v_p$, and $v_n$ are allowed to act
in the partial waves with the respective pair orbital angular momentum $l_\alpha$ up to
2, 5, and  10; the latter is not really needed for $v_n$ but is necessary for
the screened Coulomb potential  $w_{nR}$. Three-body states with the total 
angular momentum up to $J \le 20$ are included. With these cutoffs well converged
results are obtained.

The $p$-$n$ potential $v_A$ is taken to be the realistic CD Bonn potential
\cite{machleidt:01a}; the results show very little sensitivity to the choice
of $v_A$ provided it remains a realistic high-precision potential.
This is not so for the nucleon-core potentials $v_p$ and $v_n$.
The predictions are therefore obtained using several parametrizations of
the nucleon-core optical potential, namely, those of
Watson (W) \cite{watson},  Chapel Hill 89 (CH89) \cite{CH89},
Koning-Delaroche (KD) \cite{koning}, and the nonlocal (NL) potential
introduced by  Giannini and Ricco \cite{giannini} but with the parameters
readjusted in Ref.~\cite{deltuva:09b} to the experimental data for the nucleon
scattering from $\A{12}{C}$ and $\A{16}{O}$. For this reason the NL potential is not applied
to the $d+\A{7}{Be}$ reaction. Note that CH89 and KD potentials were originally
fitted to heavier nuclei $A\ge 24$ data but nowadays are often used also for light nuclei
such as isotopes of carbon or beryllium \cite{dBe12-21} 
and provide quite a reasonable description. The Watson potential was designed 
for the light $p$-shell nuclei but it is rather old and may lack accuracy. The NL parameters are 
energy-independent, while for local energy-dependent potentials they are taken at half energy of the deuteron
beam $E_d/2$ as proposed in Refs.~\cite{johnson:70a,johnson:72a}.
In the partial waves with the core-nucleon bound state the potentials
$v_p$ and $v_n$ must be real and energy-independent.
In the coordinate space they are taken to have central and
spin-orbit parts, i.e.,  
\begin{equation}  \label{eq:Vb}
v_{\alpha}(r) = -V_c f(r,R,a) + \vecg{\sigma}\cdot \mbf{L} \,
V_{so} \, \frac{2}{r} \frac{d}{dr}f(r,R,a),
\end{equation}
with $f(r,R,a) = [1+\exp((r-R)/a)]^{-1}$ and standard values for the parameters
$R = r_0 A^{1/3}$,  $r_0 = 1.25$ fm, $a=0.65$ fm, and  $V_{so}= 6.0 \, \mathrm{MeV}\cdot\mathrm{fm}^2$,
while $V_c$ is adjusted to reproduce the desired binding energy.
For all considered final-state nuclei the binding energies $\epsilon_\alpha$, the valence particle
quantum numbers, and the respective $V_c$ values are collected in Table I.
Since the excitations of the core $A$ are not taken into account, all these nuclei are single-component
states with the respective spectroscopic factor being unity.

\begin{table} [!]
\caption{\label{tab:EB}
Valence nucleon quantum numbers in the spectroscopic $nlj$ notation, the binding energies $\epsilon_\alpha$,
and the potential strengths  $V_c$ for the considered final-state nuclei.
For $\A{17}{F}(1/2^+)$ and $\A{17}{O}(1/2^+)$ the Pauli forbidden  $1s_{1/2}$ state
is projected out.}
\begin{ruledtabular}
\begin{tabular}{l*{3}{r}}
 & valence & $\epsilon_\alpha$(MeV) & $V_c$(MeV) \\ \hline
$\A{8}{B}(2^-)$    & $1p_{3/2}$ & 0.137 & 43.074 \\
$\A{13}{N}(1/2^-)$ & $1p_{1/2}$ & 1.944 & 44.361 \\
$\A{13}{C}(1/2^-)$ & $1p_{1/2}$ & 4.946 & 44.365 \\
$\A{17}{F}(5/2^+)$ & $1d_{5/2}$ & 0.600 & 52.858 \\ 
$\A{17}{F}(1/2^+)$ & $2s_{1/2}$ & 0.105 & 53.002 \\ 
$\A{17}{O}(5/2^+)$ & $1d_{5/2}$ & 4.143 & 52.813 \\ 
$\A{17}{O}(1/2^+)$ & $2s_{1/2}$ & 3.272 & 53.167  
\end{tabular}
\end{ruledtabular}
\end{table}

 The choice to fix the parameters of the local optical potential at 
$E_d/2$ was proposed in Refs.~\cite{johnson:70a,johnson:72a} and has been used widely,
but recently has been criticized
in Ref.~\cite{timofeyuk:13b} from the adiabatic distorted-wave  approximation (ADWA) point
of view, suggesting the energy that is higher by about 40 MeV. However, such a high value 
$E_d/2 + 40$ MeV at low $E_d$ is
not consistent with the  three-body  Faddeev/AGS formalism where, when integrating 
over the spectator momentum $q_\alpha$, 
the two-body subsystem energy  $E_\alpha = E - q_\alpha^2/2M_\alpha$
runs from the maximal value $E = E_d A/(A+2) - 2.225$ MeV to $-\infty$.
Although Ref.~\cite{deltuva:09a} took this aspect into account and 
used an extended Faddeev/AGS formalism allowing
optical potentials that vary with the two-body subsystem energy, the standard calculations
in the present work are performed with  optical potentials
taken at a fixed energy in order to have a single three-body  Hamiltonian and thereby 
preserve a Hamiltonian theory. For the same reason the binding potentials \eqref{eq:Vb}
are chosen energy-independent as well.
These inconsistencies and ambiguities
 arise due to the reduction of the $(A+2)$ body problem to the three-body problem.
The related uncertainties in the present results will be discussed at the end of this section.

\begin{figure}[!]
\begin{center}
\includegraphics[scale=0.82]{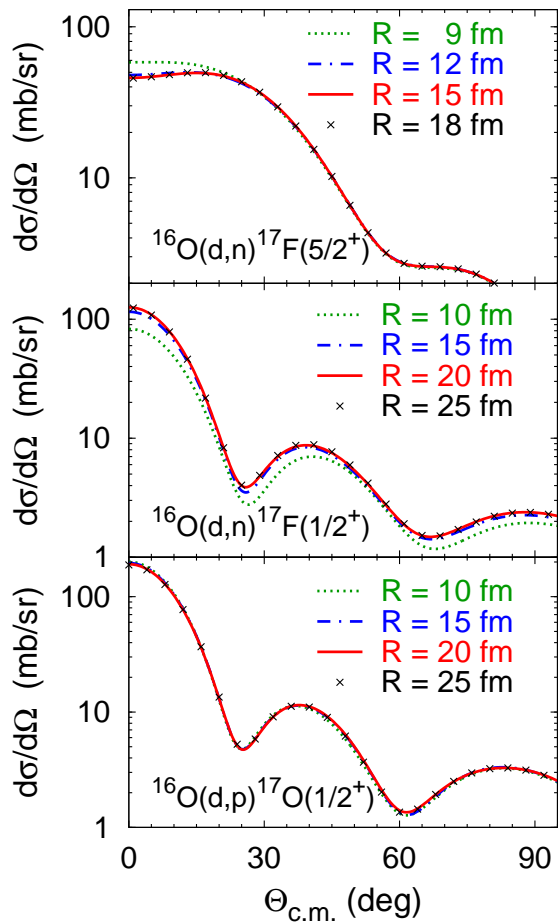}
\end{center}
\caption{\label{fig:R}  (Color online)
Differential cross section for $\A{16}{O}(d,n)\A{17}{F}$
and $\A{16}{O}(d,p)\A{17}{O}$ transfer reactions  at $E_d = 12$ MeV.
Results obtained with different values of the Coulomb screening
radius $R$ are compared. The  screening smoothness parameter $\bar{n}=8$.}
\end{figure}

To demonstrate the numerical reliability of the proposed 
calculational scheme for $(d,n)$ reactions, 
the convergence with the Coulomb screening radius $R$ is studied in Fig.~\ref{fig:R}.
As a working example the reactions with the strongest Coulomb interaction, i.e., 
 $\A{16}{O}(d,n)\A{17}{F}$, are chosen. Since $\A{17}{F}$ has ground and excited states, 
 the convergence is checked for different values
of the orbital angular momentum $l_n$ and of the binding energy $\epsilon_n$ for the bound
 proton-core pair as listed in Table I. In the  case of
the $\A{17}{F}(1/2^+) $ excited state, 
owing to its very weak binding $\epsilon_n=0.105$ MeV, a more significant sensitivity to the
Coulomb screening radius $R$ may be expected. This is indeed so as  Fig.~\ref{fig:R} shows
for the differential cross section $d\sigma/d\Omega$ as  a function of the
c.m. scattering angle $\Theta_{\cm}$ at deuteron beam energy $E_d = 12$ MeV.
While for the transfer to the ground state $\A{17}{F}(5/2^+)$ the $R$-independence
is established at $R > 12$ fm, the same level of convergence for the transfer to 
the  excited state $\A{17}{F}(1/2^+)$ is reached only at $R \ge 20$ fm.
For comparison, $d\sigma/d\Omega$  is presented also for the $\A{16}{O}(d,p)\A{17}{O}(1/2^+)$ 
reaction where the final-channel two-particle bound state is 
Coulomb-free but the two-cluster scattering
state has to be renormalized. The convergence with  $R$ is considerably faster in this case.
Thus, compared to $(d,p)$ the calculation of $(d,n)$ reactions is more demanding
in terms of the screening radius but, nevertheless, well-converged results are obtained.

In the following the physics results for $\A{7}{Be}(d,n)\A{8}{B}$,
$\A{12}{C}(d,n)\A{13}{N}$, and  $\A{16}{O}(d,n)\A{17}{F}$ reactions are 
presented and compared with the experimental data.

\begin{figure}[!]
\begin{center}
\includegraphics[scale=0.7]{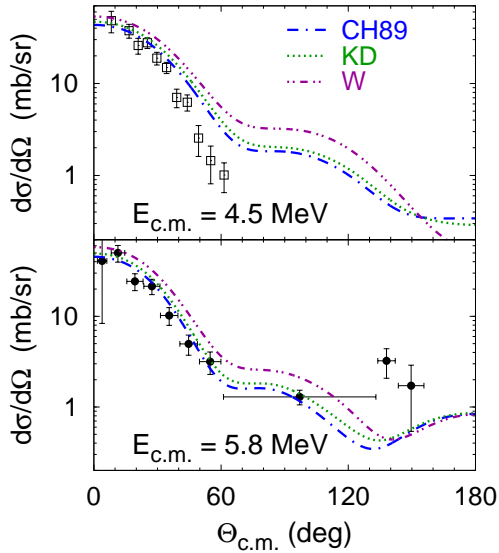}
\end{center}
\caption{\label{fig:Be}  (Color online)
Differential cross section for the $\A{7}{Be}(d,n)\A{8}{B}$ transfer reaction
  at $E_\cm =  4.5$ and 5.8 MeV. Predictions obtained using 
Chappel Hill 89 (dashed-dotted curves), Koning-Delaroche (dotted curves),
and Watson (dashed-double-dotted curves) optical potentials are compared
with the experimental data from Refs.~\cite{PhysRevLett.77.611} (5.8 MeV) and
\cite{PhysRevC.73.015808} (4.5 MeV).}
\end{figure}

I start with the $\A{7}{Be}(d,n)\A{8}{B}$ reaction where the experimental
data are rather scarce and a bit contradictory \cite{PhysRevLett.77.611,PhysRevC.73.015808}.
Differential cross section results 
at $d+\A{7}{Be}$  kinetic c.m. energy of 4.5 and 5.8 MeV 
obtained using CH89, KD, and W  optical potential models  are given in Fig.~\ref{fig:Be}.
The predictions of CH89 are closest to the data but one should keep in mind that this potential
as well as KD is not fitted to the nucleon-$\A{7}{Be}$ data.
All calculations reproduce reasonably the angular shape of $d\sigma/d\Omega$ but
overpredict its magnitude. This is not unexpected given the existence of low-energy
excitations of the $\A{7}{Be}$ core that are not taken into account in the present calculations. 
Relying on the analogy with the $\A{10}{Be}(d,p)\A{11}{Be}$ reactions \cite{deltuva:13d}
one may expect  that including the  $\A{7}{Be}$ core excitation would lead to
the reduction of the transfer cross section, at least at forward angles.

\begin{figure}[!]
\begin{center}
\includegraphics[scale=0.62]{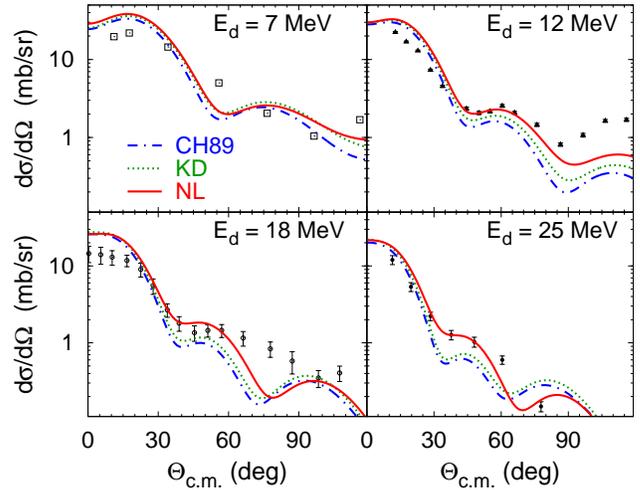}
\end{center}
\caption{\label{fig:C}  (Color online)
Differential cross section for the  $\A{12}{C}(d,n)\A{13}{N}$  reaction
at 7, 12, 18, and 25 MeV deuteron energy. Predictions obtained using the
Chappel Hill 89 (dashed-dotted curves), Koning-Delaroche (dotted curves),
and nonlocal (solid curves) optical potentials are compared
with the experimental data from Refs.
\cite{J.NP_A.414.67.1984} (7 MeV), \cite{J.NP_A.172.469.197109}  (12 MeV),
\cite{J.SNP.48.403.1988} (18 MeV),
and \cite{J.JP_G.38.075201.2011} (25 MeV).}
\end{figure}

\begin{figure}[!]
\begin{center}
\includegraphics[scale=0.62]{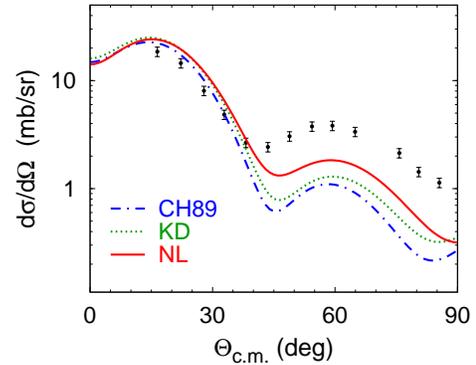}
\end{center}
\caption{\label{fig:Cdp}  (Color online)
Differential cross section for the  $\A{12}{C}(d,p)\A{13}{C}$  reaction
at 12 MeV deuteron energy. Curves are as in Fig.~\ref{fig:C},
and the  experimental data are from Ref.~\cite{J.NP_A.477.77.1988}.
}
\end{figure}

Next I show in Fig.~\ref{fig:C} the differential cross section for 
the $\A{12}{C}(d,n)\A{13}{N}$ transfer reaction at $E_d = 7$, 12, 18, and 25 MeV.
The used optical potentials are  CH89, KD, and NL. For all of them the  angular shape 
 fails in accounting for the experimental data: the data points from
Refs.~\cite{J.NP_A.414.67.1984,J.NP_A.172.469.197109,J.SNP.48.403.1988,J.JP_G.38.075201.2011}
are overestimated 
at forward angles $\Theta_{\cm} < 30^\circ$ and underestimated at large angles
 $\Theta_{\cm} > 90^\circ$. This discrepancy is comparable with the one in the
$\A{12}{C}(d,p)\A{13}{C}$ reaction \cite{J.NP_A.477.77.1988}, displayed  in  Fig.~\ref{fig:Cdp}.
A somehow similar discrepancy, i.e., overprediction at small angles and underprediction
of the data at larger angles,
 was observed also at higher energies in the $\A{12}{C}(d,p)\A{13}{C}$ reaction,
but only for the $\A{13}{C}$ ground state $1/2^-$; the agreement for the transfer to
 $\A{13}{C}$ excited states $1/2^+$ and  $5/2^+$ was significantly better, especially when
using the nonlocal optical potential \cite{deltuva:09b}. In this view the failure of 
the calculations to reproduce the $\A{12}{C}(d,n)\A{13}{N}$ data is not unexpected.

\begin{figure*}[!]
\begin{center}
\includegraphics[scale=0.88]{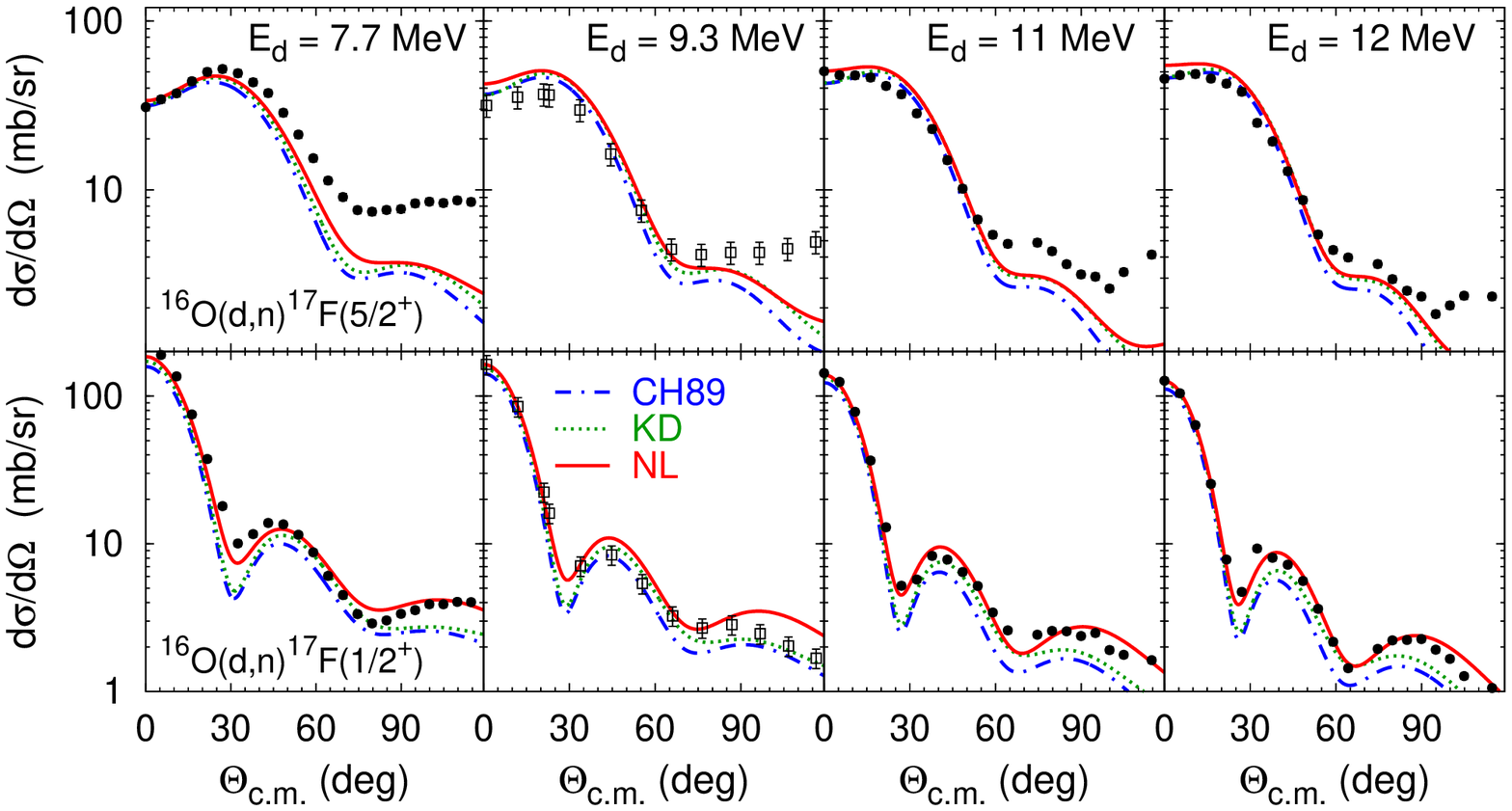}
\end{center}
\caption{\label{fig:O}  (Color online)
Differential cross section for   $\A{16}{O}(d,n)\A{17}{F}$  reactions
leading to the $\A{17}{F}$ ground state $5/2^+$ (top) and 
excited state  $1/2^+$ (bottom)
at the deuteron energy  $E_d = 7.7$, 9.3, 11, and 12 MeV.
Curves are as in Fig.~\ref{fig:C}, and the experimental  
data are from Refs.~\cite{Thornton1969531} (9.3 MeV) and \cite{Oliver1969567}
(other energies).}
\end{figure*}
\begin{figure*}[!]
\begin{center}
\includegraphics[scale=0.8]{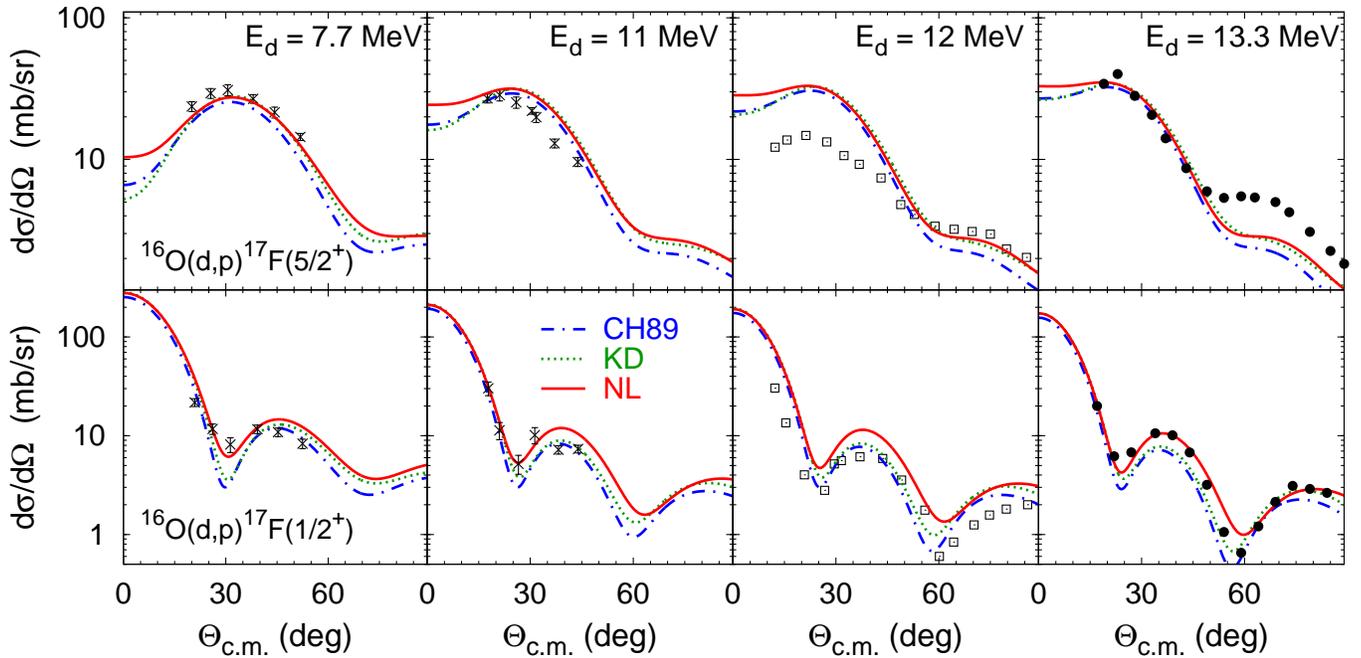}
\end{center}
\caption{\label{fig:Odp}  (Color online)
Differential cross section for   $\A{16}{O}(d,p)\A{17}{O}$  reactions
leading to the $\A{17}{O}$ ground state $5/2^+$ (top) and 
excited state  $1/2^+$ (bottom)
at the deuteron energy  $E_d = 7.7$, 11, 12, and 13.3 MeV.
Curves are as in Fig.~\ref{fig:C}, and the experimental  
data are from Refs.~\cite{J.NP_A.112.76.1968} (7.7 and 11 MeV),
\cite{J.NP_A.97.541.1967} (12 MeV), and \cite{J.NP_A.188.164.1972} (13.3 MeV).}
\end{figure*}

Finally I consider the $\A{16}{O}(d,n)\A{17}{F}$ reactions.
Unfortunately, the available experimental data for the 
angular distribution of the cross section,
to the best of my knowledge, are limited to low energies
$E_d \le 12$ MeV. The theoretical predictions
based on CH89, KD, and NL optical potentials 
at  $E_d = 7.7$, 9.3, 11, and 12 MeV are presented in  Fig.~\ref{fig:O}
and compared with the experimental data from Refs.~\cite{Thornton1969531,Oliver1969567}.
For the transfer to the $\A{17}{F}$ ground state $5/2^+$
the theoretical results follow the data \cite{Oliver1969567}
up to $\Theta_{\cm} = 30^\circ$ 
at $E_d = 7.7$ MeV, up to $ 60^\circ$ at  $E_d = 11$ MeV, 
and up to $ 90^\circ$ at  $E_d = 12$ MeV
but underpredict at larger angles. This discrepancy is most sizable 
at  $E_d = 7.7$ MeV, reaching a factor of 2, possibly
due to the compound-nucleus reaction mechanism that is expected to become  relevant
with decreasing energy.
The data from  Ref.~\cite{Thornton1969531} is slightly overestimated 
at small angles possibly indicating  some inconsistency between
the two sets \cite{Thornton1969531,Oliver1969567}.
The agreement is better for the transfer to the $\A{17}{F}$ excited state $1/2^+$.
While  CH89 and KD results slightly underestimate the data, the nonlocal
potential NL reproduces well the experimental data in the whole
angular regime  up to $\Theta_{\cm} = 120^\circ$. This possibly indicates 
that a simple proton plus core model for the $\A{17}{F}$ excited state $1/2^+$
is adequate, and supports the conjecture of Ref.~\cite{PhysRevC.59.1211}
on this state being one-proton halo. This is not unexpected given
the very weak binding of the $\A{17}{F}(1/2^+)$ nucleus.

The results for  $\A{16}{O}(d,p)\A{17}{O}$ reactions 
at  $E_d = 7.7$, 11, 12, and 13.3 MeV are presented  in  Fig.~\ref{fig:Odp}.
The agreement between the theoretical predictions and experimental
data \cite{J.NP_A.112.76.1968,J.NP_A.97.541.1967,J.NP_A.188.164.1972}
is almost as good as in the case of  Fig.~\ref{fig:O}, the exception
being the $E_d=12$ MeV data from Ref.~\cite{J.NP_A.97.541.1967} whose normalization seems
to be inconsistent with the other sets.
It is noteworthy that the nonlocal potential NL whose parameters, 
in contrast to the local potentials CH89 and KD,
are independent of the energy,  provides a successful description
of $\A{16}{O}(d,p)\A{17}{O}$ and  $\A{16}{O}(d,n)\A{17}{F}$ reactions
over a broad range of energies (see Ref.~\cite{deltuva:09b} for $(d,p)$ reactions
at higher energies).

\begin{figure}[!]
\begin{center}
\includegraphics[scale=0.75]{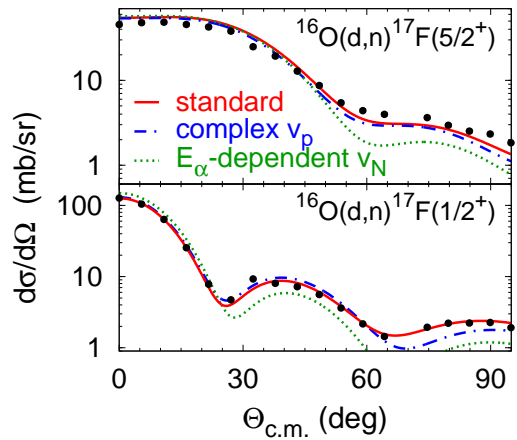}
\end{center}
\caption{\label{fig:V}  (Color online)
Differential cross section for   $\A{16}{O}(d,n)\A{17}{F}$  reactions
at   $E_d = 12$ MeV. Results of standard calculations using the NL potential
(solid curves) are compared with the results using the complex $nA$ potential
in all partial waves (dashed-dotted curves) and with the results using 
energy-dependent $nA$ and $pA$ potentials in the  partial waves with bound states
(dotted curves). The experimental  
data are from Ref.~\cite{Oliver1969567}.}
\end{figure}

To study the sensitivity of the results to some ambiguities in the chosen Hamiltonian,
two additional types of calculations are presented in Fig.~\ref{fig:V}
for $\A{16}{O}(d,n)\A{17}{F}$ reactions at $E_d=12$ MeV.
First, the real binding potential in the $nA$ partial waves with bound states is replaced
by the complex NL potential, i.e., the 
$\A{17}{O}$ bound states listed in Table I are not supported, and the 
$(d,p)$ reactions are not taking place. Nevertheless, the results for the
$(d,n)$ reactions remain almost unaltered, except at large angles
 $\Theta_{\cm} > 60^\circ$ for the transfer to the $\A{17}{F}$ excited state $1/2^+$.
Second, both $nA$ and $pA$ potentials in the  partial waves with bound states
are kept real at negative two-body subsystem energies $E_\alpha$ but are taken 
to be complex NL potentials for  $E_\alpha > 0$. Such an abrupt
change has been chosen to maximize the possible effect.
Although the AGS equations can be solved with such potentials \cite{deltuva:09a},
they do not correspond to a definite Hamiltonian, so the conclusions from 
such calculations have to be taken with care.
The effect of the allowed energy-dependence is small for 
the transfer to the $\A{17}{F}$ ground state $5/2^+$ up to  $\Theta_{\cm} < 45^\circ$
 but is of moderate size for the transfer to the $\A{17}{F}$ excited state $1/2^+$,
even at forward angles. It appears to be correlated with the
binding energy. A possible explanation is that the two-body $t$-matrix pole, located at 
$E_\alpha = - \epsilon_\alpha$ affects also the $E_\alpha > 0$ region if the binding
is weak as in the case of  the $\A{17}{F}$ excited state $1/2^+$.
In addition to being non-Hamiltonian,
the energy-dependent potential destroys the consistency  between
the bound state features and threshold behavior that is important 
in low-energy reactions. The standard calculations of this paper should
therefore be considered as more adequate.

\section{Summary \label{sec:4}}

Proton transfer reactions in the
deuteron-nucleus collisions were described in a three-body model
for the proton + neutron + nuclear core system. The framework of 
 exact integral equations for three-body transition operators
as proposed by Alt, Grassberger, and Sandhas was used; they were
solved in the momentum-space partial-wave representation.
 The Coulomb interaction between the proton and the core was
included via the screening and renormalization method. 
The differences  relative to the calculation of neutron transfer reactions
were pointed out. When the final-state nucleus is weakly bound, e.g.,
 the $\A{17}{F}$ in the excited state $1/2^+$, the convergence 
with the screening radius becomes slower for the $(d,n)$ reactions
as compared to  $(d,p)$, thereby making the numerical $(d,n)$ calculations
more demanding. Nevertheless, well converged results were obtained
for the differential cross section of $\A{7}{Be}(d,n)\A{8}{B}$,
$\A{12}{C}(d,n)\A{13}{N}$, and  $\A{16}{O}(d,n)\A{17}{F}$ reactions.

Realistic CD Bonn potential was used for the interaction between nucleons.
In  partial waves with the nucleon-core bound states the potential strength was adjusted to reproduce
the experimental binding energies while in the other partial waves one of the four optical potentials,
i.e.,  Watson,  Chapel Hill 89, Koning-Delaroche, and the nonlocal one
by  Giannini and Ricco, was used.
For the $\A{7}{Be}(d,n)\A{8}{B}$ reaction
the calculations reproduced reasonably the angular shape of the  
differential cross section data  but
overpredicted its magnitude, presumably due to the neglection of the  $\A{7}{Be}$ core excitations.
A disagreement with the experimental data in the angular distribution of 
$d\sigma/d\Omega$ was observed in  the $\A{12}{C}(d,n)\A{13}{N}$ reaction,
similar to earlier findings in the corresponding $(d,p)$ reaction.
The description of the  $\A{16}{O}(d,n)\A{17}{F}$ reactions was
 quite successful, especially 
for the  transfer to the $\A{17}{F}$ excited state $1/2^+$
using the energy-independent nonlocal optical potential that
 reasonably reproduced also the  $\A{16}{O}(d,p)\A{17}{O}$ data over a
broader energy range. 

\vspace{3mm}

This work was supported by the Research Council of Lithuania under 
contract No.~MIP-094/2015.

\vspace{1mm}



\end{document}